# Experimental Demonstration of Quantum Entanglement Between Frequency-Nondegenerate Optical Twin Beams

Xiaolong Su, Aihong Tan, Xiaojun Jia*, Qing Pan, Changde Xie, Kunchi Peng

State Key Laboratory of Quantum Optics and Quantum Optics Devices,
Institute of Opto-Electronics, Shanxi University, Taiyuan 030006, China

**Abstract**: The quantum entanglement of amplitude and phase quadratures between two intense optical beams with the total intensity of 22mW and the frequency difference of 1nm, which are produced from an optical parametric oscillator operating above threshold, is experimentally demonstrated with two sets of unbalanced Mach-Zehnder interferometers. The measured quantum correlations of intensity and phase are in reasonable agreement with the results calculated based on a semi-classical analysis of the noise characteristics given by C. Fabre et al.

In recent years quantum information with continuous variables (CV) has been extensively investigated[1]. Continuous-wave (CW) degenerate or non-degenerate optical parametric amplifiers (DOPA or NOPA) operating below threshold have been one of the best schemes producing entangled states of light. The existence of entanglement between amplitude and phase quadratures of intense signal and idler modes (named as twin beams) produced from optical parametric oscillator (OPO) operating above threshold was theoretically predicted by Reid and Drummond in 1988 firstly and then was successively analyzed with detailed theoretical calculations[2-6]. Especially, C. Fabre et al. presented very useful expressions for calculating quantum correlations of amplitude and phase components between output twin beams from a non-degenerate OPO (NOPO) above threshold by means of a semi-classical analysis of the noise characteristics[4]. The noise spectra of the intensity difference ($S_I(f)$) and phase sum ($S_P(f)$) of twin beams are expressed respectively by [4]:

$$S_I(f) = S_0(1 - \frac{\eta\xi}{1+(f/B)^2}) \tag{1}$$

$$S_P(f) = S_0(1 - \frac{\eta\xi}{\sigma^2+(f/B)^2}) \tag{2}$$

where $f$ is the noise frequency, $S_0$ is the shot noise limit (SNL), $B$ and $\xi = T/(T+\delta)$ are the cavity bandwidth and the output coupling efficiency of NOPO respectively ($T$ - the transmission coefficient of the output coupling mirror; $\delta$ – extra intracavity losses), $\eta$ is the detection efficiency, $\sigma = \sqrt{P/P_0}$ is the pump parameter ($P$ - the pump power, $P_0$ - the threshold pump power of NOPO). The intensity difference quantum correlations of twin beams were experimentally measured with self-homodyne detectors by different groups and were effectively applied[7-12]. However, the phase correlation of the twin beams was not observed for a long time due to the technic difficulty of measuring phase noise of twin beams with non-degenerate frequencies. Up to very recent date, Laurat et al. forced the NOPO to oscillate in a strict frequency-degenerate situation by inserting a /4 plate inside the NOPO and observed the phase-sum variance of 3dB above the SNL[13]. Latter, the phase correlation of 0.8dB below the SNL ($\Delta^2 \hat{q}_+ = 0.82$) between twin beams with different frequencies from NOPO for the pump power about 4% above threshold was measured by A. S. Villar et al. by means of scanning a pair of tunable ring analysis cavity[14]. Almost at a parallel period we were also devoting our efforts to measure the quantum entanglement of twin beams from NOPO above threshold. The measurement scheme used by us is basically same with that presented by O. Glockl et al. in Ref.[15], where they performed sub-shot-noise measurement of the phase quadratures of intense pulsed light[16]. Considering that the phase correlation will be significantly affected by the phase fluctuation of the pump laser[6] and the restricted condition deducing Eqs.(1) and (2) in Ref.[4] requires the finesse of the NOPO cavity for the pump laser much lower than that for the twin beams, in our design the ratio of the cavity finesses for the pump and the twin beams is 16/164 which is much smaller than that in Refs.[13] and [14]. Due to the lower finesse the resonant peak of the pump laser in the cavity is relatively flat and thus the threshold power is higher (~120mW).

At first, using a pair of Mach-Zehnder (M-Z) interferometers with unbalanced arm-lengths we detected the amplitude and phase noise of signal and idler output fields from a NOPO above threshold at a certain analysis frequency (20MHz), respectively. Then, the quantum correlations were denoted by the noise levels of the intensity difference and the phase sum of the photocurrents measured by two unbalanced interferometers. For a unbalanced M-Z interferometer, when the relative optical phase shift ($\varphi$) between two optical fields passing through the short arm of length L and through the long arm of length L+$\Delta$L is adjusted to $\varphi=\pi/2+2k\pi$(k - integer), and at the same time the phase shift ($\theta$) of the spectral component of radio frequency (rf) fluctuations at a sideband frequency ($\Omega =2\pi f$) is controlled to $\theta=\pi$, the fluctuations of the sum and the difference photocurrents of two output fields from M-Z interferometer in frequency space are proportional to the vacuum noise level and the spectral component of the phase quadrature of the initial fields, respectively[15]. Therefore, for measuring the phase fluctuation of input field the length difference $\Delta L$ between the short and long arms should be taken as $\Delta L = c\pi/\Omega$ (c - the speed of light) to meet the condition of $\theta = \Omega \Delta L/c = \pi$.

The experimental system for the entanglement measurements is depicted in Fig.1. The NOPO consists of a 10mm long α-cut type-II KTP ($KTiOPO_4$, potassium titanyl phosphate) crystal and a concave mirror of 30mm curvature. The front face of KTP is coated to be used as the input coupler of the pump laser at 540nm, which is produced from a home-made frequency-doubled and frequency-stabilized Nd:YAP/KTP (Nd-doped $YAlO_3$ perovskite/KTP) laser[17]. The concave mirror with a transmission of 3.2% at 1080nm wavelength is the output coupler of the twin beams. The mirror is mounted on a piezo-electric transducer (PZT) for locking actively the cavity length of NOPO on resonance with the pump laser. The measured cavity finesses for 1080nm is ~164 and the total intracavity extra losses is about 0.6%, thus the output coupling efficiency ξ equals to 84%. The twin beams with cross polarized directions, are separated by a polarizing beam splitter (PBS1) firstly and then each of them is directed into a unbalanced M-Z interferometer. The 50/50 beamsplitters, BS1 and BS2, are the output mirrors of the two interferometers, respectively. The output optical fields are detected by a balanced detection system consisting of high-efficiency photodiodes D1 and D2 (D3 and D4). The input beamsplitter of the interferometer is made of a polarizing-beam-splitter PBS2 (PBS3) and a λ/2 wave plate P1 (P2). Rotating the polarization orientation of P1 (P2) we can conveniently switch between phase and amplitude measurements[15]. In our system, the distance difference of two arms $\Delta L$ is 7.5m which matches the analysis frequency of 20MHz to make $\theta=\pi$. The difference of the dc photocurrents of D1 and D2 (D3 and D4) serves as the error signal and is fed back onto the PZT mounted on one of mirrors of the interferometer to stabilize the relative optical phase between two arms at $\varphi=\pi/2+2k\pi$. A set of optical lenses M1 (M2) in the interferometer is used for the mode-matching between two beams from short and long arms on BS1 (BS2). The variances of the output photocurrent fluctuations from two interferometers were recorded with a pair of spectrum analyzers (SA), and then the correlation variances of amplitude and phase quadratures between the twin beams were denoted by the difference and sum of the corresponding photocurrents from each interferometer. For demonstrating the quantum entanglement of twin beams, the pump power (P) of NOPO is kept at 230mW which is 110mW higher than the oscillation threshold power ($P_0$=120mW) during whole measurements. Although the phase correlation should be better for the pump power close to the threshold (Eq.(2)), we found that the NOPO with lower finesse for the pump laser could stably operate only in the case using higher pump power at about twice of the threshold and the output was very unstable when the pump power approached the threshold. We consider the reason is that the multiple modes can simultaneously oscillate in the NOPO with the low finesse of the pump laser due to its flat resonance peak thus the mode-competition and mode-hoping must induce the instability. When the pump power is increased, once the oscillating of a twin-beam mode dominates absolutely in the NOPO, the output will be stable. Under the operating condition the detected output power of the twin beams is 22mW, which are the experimentally obtained most intense entangled beams so far to the best of our knowledge. The measured wavelengths of the signal and idler beams are 1080.215nm and 1079.130nm, respectively ($\Delta\lambda$=1.085nm).

The difference of two amplitude noise powers and the sum of two phase noise powers, are given in Fig.2 (a)ii and (b)ii, respectively. The correlation variances of the intensity difference and the phase sum between the twin beams are below the SNL of twin beams (trace i in (a) and (b)) of 1.25±0.06dB and 0.60±0.07dB, respectively. After accounting for the influences of the electronics noise of ~3.9dB below the SNL the correlations of amplitude and phase quadratures should be about 2.58dB and 1.05dB below the SNL, respectively. The sum of the correlation variances of the amplitude (X) and phase (Y) quadratures of the twin beams equals to $<\delta(\frac{X_1-X_2}{\sqrt{2}})^2> + <\delta(\frac{Y_1+Y_2}{\sqrt{2}})^2> = 1.332$, which is less than the SNL normalized to 1 for each combination of quadratures (The SNL of whole twin beams is 2)[14]. Thus the quantum entanglement between the signal and idler optical beams with non-degenerate frequencies generated from a NOPO above threshold is experimentally proved according to the inseparability criterion proposed by Duan et al.[18].

Fig.3 gives the normalized noise power spectra of intensity difference (i) and phase sum (ii) calculated from Eqs.(1) and (2) using the parameters of the real experimental system (ξ=0.84, B=24.7MHz, σ=1.38, η=0.88 ).The normalized correlation variances measured at 20MHz (Fig.2) are marked on Fig.3 with a star(  ) for amplitude and a dot(•) for phase. The experimental values are in reasonable agreement with the calculated results. The measured phase correlation noise is higher of 0.08 than that of theoretical calculation, that is because the imperfect mode matching efficiency between the two beams from the short and long arms of the interferometer and the influence of the phase fluctuation of the pump field are not involved in Eq.(2). The measured mode-matching efficiency is about 90% which introduces the extra noise of ~0.03 on the phase quadratures[19]. The remained excessive noise of ~0.05 may come from the pump fluctuation, perhaps.

In conclusion, we have experimentally demonstrated the quantum entanglement of frequency non-degenerate twin beams produced from a CW NOPO operating above threshold using a pair of unbalanced M-Z interferometers. The twin beams with high mean intensities of 22mW are easier to be manipulated and utilized, therefore the presented scheme can be conveniently applied in future quantum communication of CV.

Acknowledgements: This work was supported by the National Natural Science Foundation of China (No.60238010, 60378014) and the Major State Basic Research Project of China (No.2001CB309304). X. Jia's email address is jiaxj@sxu.edu.cn

List of figure captions

1. Schematic of the experimental setup. Nd:YAP/KTP: laser source; NOPO: nondegenerate optical parametric oscillator; $PBS_{1-3}$: polarizing-optical-beamsplitter; $P_{1-2}$: $\lambda/2$ wave plate; $BS_{1-2}$: 50/50 optical beamsplitter; $M_{1-2}$: Mode matching lense; +/-: positive/negative power combiner; $D_{1-4}$: photodiode detector (ETX500 InGaAs); SA: spectrum analyzer

2. The amplitude (a) and phase (b) correlation noise powers of twin beams at 20MHz. (i): SNL; (ii): correlation noise; (iii): ENL. The measurement parameters of SA (Spectrum Analyzer): RBW (Resolution Band Width)-10kHz; VBW (Video Band Width)-30Hz.

3. The normalized noise power spectra of (i) amplitude difference and (ii) phase sum calculated from theoretical analysis. The star (  ) and the dot (•) correspond to the experimental values for amplitude (0.55 corresponding to 2.58dB below the SNL) and phase (0.78 corresponding to 1.05dB below the SNL) correlation respectively.

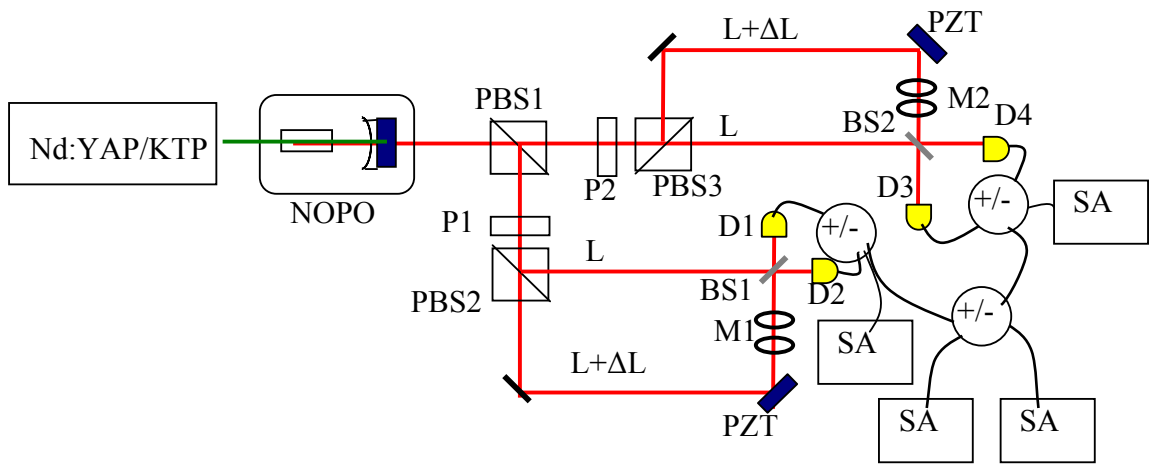

Fig.1

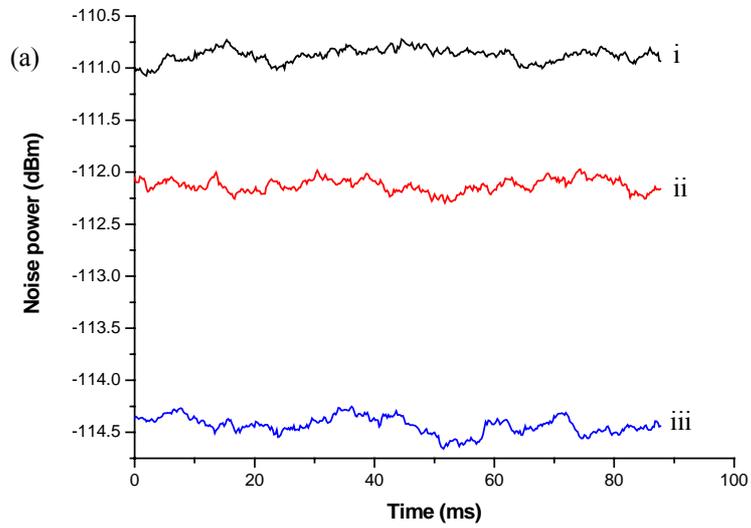

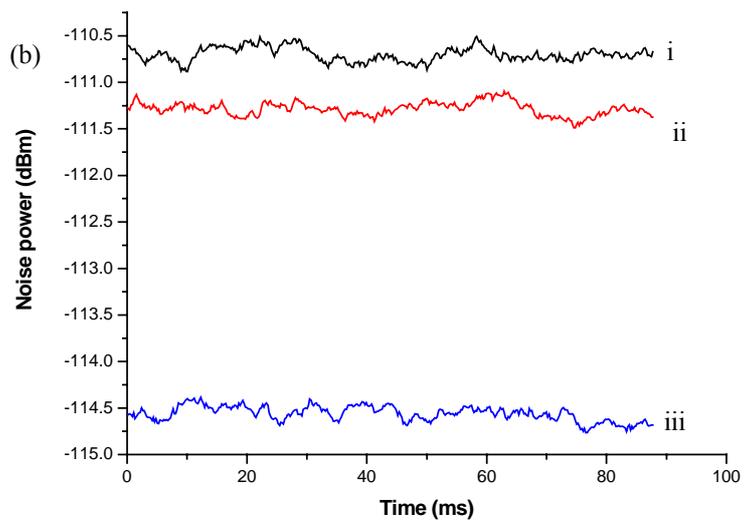

Fig.2

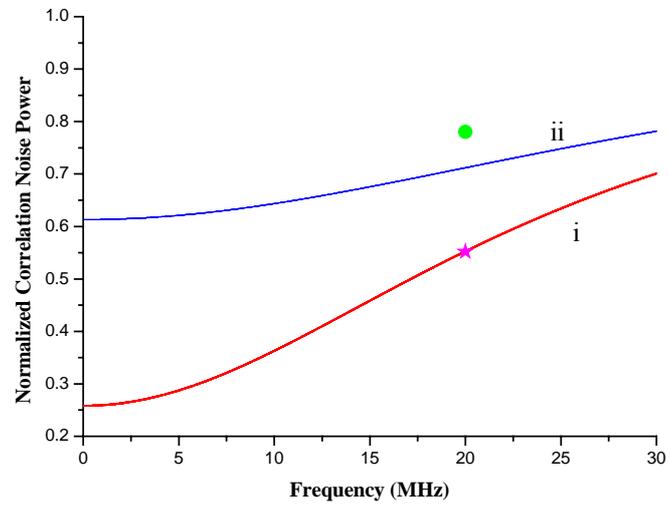

Fig.3